\documentclass{cpbtex}

\usepackage{multirow}
\usepackage{graphicx}

\begin{document}
%\begin{CJK*}{GBK}{song}

\title{Robustness of ferromagnetism in van der Waals magnet Fe$_3$GeTe$_2$ to hydrostatic pressure}

\author{Yonglin Wang$^{1}$, Xu-Tao Zeng$^{1}$,  Bo Li$^{1}$, Cheng Su$^{1}$,Takanori Hattori$^{2}$,  Xian-Lei Sheng$^{1}$, \\
and Wentao Jin$^{1,}$ \thanks{Corresponding author. E-mail: wtjin@buaa.edu.cn} 
%中文名：王涌霖，曾旭涛，李博，宿程，Takanori Hattori，胜献雷，金文涛%
\\
\\
$^{1}${School of Physics, Beihang University, Beijing 100191, China}\\  % The line break was forced via \\
$^{2}${J-PARC Center, Japan Atomic Energy Agency, Tokai, Ibaraki 319-1195, Japan}\\ % The line break was forced via \\
} 

% 1. For Chinese authors, the name in Chinese characters should also be given. For example, Gang Liu(����), Xiao-Ming Li(������)
% 2. Please ensure that every author approves the submission of the manuscript
% 3. Abbreviations should not be used in the affiliations

\date{\today}
\maketitle

\begin{abstract}
Two-dimensional van der Waals ferromagnet Fe$_3$GeTe$_2$ (FGT) holds a great potential for applications in spintronic devices, due to its high Curie temperature, easy tunability, and excellent structural stability in air. Theoretical studies have shown that pressure, as an external parameter, significantly affects its ferromagnetic properties. %It can even increase the Curie temperature to room temperature under certain circumstances\cite{21,22}. 
In this study, we have performed comprehensive high-pressure neutron powder diffraction (NPD) experiments on FGT up to 5 GPa, to investigate the evolution of its structural and magnetic properties with hydrostatic pressure. The NPD data clearly reveal the robustness of the ferromagnetism in FGT, despite of an apparent suppression by hydrostatic pressure. As the pressure increases from 0 to 5 GPa, the Curie temperature is found to decrease monotonically from 225(5) K to 175(5) K, together with a dramatically suppressed ordered moment of Fe,  which is well supported by the first-principles calculations. Although no pressure-driven structural phase transition is observed up to 5 GPa,  quantitative analysis on the changes of bond lengths and bond angles indicate a significant modification of the exchange interactions, which accounts for the pressure-induced suppression of the ferromagnetism in FGT.
\end{abstract}



\section{Introduction}
Since the successful isolation of monolayer graphene in 2004 \cite{1}, research into two-dimensional (2D) van der Waals (vdW) materials has surged, motivated by their atomically thin structure and exceptional physical properties \cite{2,3,4,5,6,7}. The Mermin-Wagner theorem \cite{8}, which predicts the absence of long-range magnetic order in isotropic 2D systems, was challenged by the discoveries of CrI$_3$ and Cr$_2$Ge$_2$Te$_6$ in 2017 \cite{9,10}. Both materials display intrinsic ferromagnetism at the monolayer level, owing to significant magnetic anisotropies that counteracts thermal fluctuations. This breakthrough has led to the predictions and synthesis of a variety of 2D vdW materials, those exhibiting unique properties such as tunneling magnetoresistance, quantum spin Hall effects, and spin-orbit torque effects\cite{11,12,13,14}. These materials are reshaping the future of spintronics, logic circuits, and magnetic storage technologies \cite{15,16,17}. Moreover, 2D vdW materials provide an ideal platform for investigating and manipulating various physical properties using external stimuli such as light, pressure, magnetic and electric fields \cite{5,18,19}.

Fe$_{3-\delta}$GeTe$_2$ (FGT), a famous 2D vdW ferromagnet, is anticipated to be applicable in spintronics due to its high Curie temperature ($T\rm_{C}$), which typically ranges from approximately 150 K to 220 K depending on the Fe vacancy ($\delta$). It features a sandwich-like structure, in which the Fe$_{3-\delta}$Ge slabs consisting of Fe(1)-Fe(1) pairs across a hexagonal Fe(2)-Ge network are clamped between two Te layers, with a possible deficiency ($\delta$) at the Fe(2) site that is largely correlated with the value of $T\rm_{C}$ \cite{20}. Notably, 2D vdW materials like FGT, CrI$_3$, and Cr$_2$Ge$_2$Te$_6$ exhibit strong covalent bonding within the layers and weak van der Waals forces between the layers \cite{9,10}. These materials possess large interlayer spacings, resulting in limited interlayer charge transfers and weak interlayer interactions. 

Below $T\rm_{C}$, stoichiometric Fe$_3$GeTe$_2$ is ferromagnetically ordered with both the Fe(1) and Fe(2) moments aligned along the $c$ axis \cite{20}. A large number of studies using external perturbations including light, gate voltage and strain have been conducted on FGT to tune its magnetic properties \cite{21,22,23}. Furthermore, theoretical studies suggest that pressure, as a relatively clean and homogeneous tuning parameter, also significantly impacts the ferromagnetism of FGT.  Under a uniaxial or biaxial strain, which is a 1D or 2D pressure, respectively, the $T\rm_{C}$ can even be elevated to room temperature\cite{24,25}, which opens up possibilities for practical spintronic applications. This hypothesis was later supported experimentally, showing that ferromagnetism is strengthened when a uniaxial strain is applied with the $ab$ plane\cite{26}. In contrast, under a three-dimensional (3D) hydrostatic pressure, FGT exhibits an effective suppression of its ferromagnetism \cite{27,28,29,30,31,32,33}. %Under hydrostatic pressure, the interlayer spacings can be effectively reduced, enhancing interlayer coupling and enabling in situ investigations of related physical property changes. 
However, these high-pressure experimental studies have been mainly focused on macroscopic characterizations such as magnetometry and transport measurements, while microscopic magnetic probes under a hydrostatic pressure are still quite limited, to the best of our knowledge. The only available example is a synchrotron M{\"o}ssbauer source spectroscopic study, which is however a local probe with the x-ray beam size of a few micrometers only \cite{28}. To gain deeper insights into the suppressive effect of hydrostatic pressure on the ferromagnetism of FGT, a neutron diffraction measurement, as a typical microscopic and bulk probe to its underlying magnetic structure, is quite necessary.%daozheli

In this study, we have conducted comprehensive high-pressure neutron powder diffraction (NPD) experiments on an almost stoichiometric polycrystalline sample of FGT up to 5 GPa, to investigate the evolution of the structure and ferromagnetism with the hydrostatic pressure. Although both the Curie temperature $T\rm_{C}$ and the ordered magnetic moment of Fe are significantly suppressed by the pressure, the ferromagnetism is found to be quite robust with an expected quantum critical point (QCP) around $P\rm_{C}$ = 22(2) GPa. The shrinkage of the Fe(1)-Fe(1)/Fe(2) distances and the progressive deviation from 90° of the Fe(1)-Ge/Te-Fe(1) bond angles with increasing pressure might be responsible for the suppressive effect on the ferromagnetism.

\section{\label{sec:level1}Materials and methods\protect\\}

Polycrystalline samples of FGT were synthesized through the standard solid-state reaction method. Stoichiometric amounts of Fe (99.99$\%$), Ge (99.99$\%$), and Te (99.99$\%$) powders were mixed, ground inside an argon-filled glovebox and transferred into a quartz ampoule. The quartz ampoule was evacuated, held at 675 °C for 10 days and cooled down to room temperature. The resulting black powders were examined by x-ray powder diffraction (XRD). The XRD data were collected using a Bruker D8 ADVANCE diffractometer in Bragg-Brentano geometry with Cu K$\alpha$ radiation ($\lambda$ = 1.5406 \AA). The dc magnetization of the polycrystalline samples in the temperature range from 120 to 300 K was measured using a Quantum Design Magnetic Property Measurement System (MPMS),  utilizing both the zero-field-cooling (ZFC) and field-cooling (FC) modes with an applied magnetic field of 100 Oe.  

Pressure-dependent NPD experiment on FGT was performed on the high-pressure neutron diffractometer PLANET (BL11) at the Materials and Life Science Experimental Facility (MLF) of J-PARC, Ibaraki, Japan, which runs in a time-of-flight (TOF) mode \cite{34}. A hydrostatic pressure up to 5 GPa was generated using a low-temperature MITO system, which can access a base temperature of 77 K\cite{35}. The power sample was loaded in the TiZr gasket along with a Pb pressure marker and the pressure transmitting medium of 4:1 deuterated methanol-ethanol mixture. The applied pressure was estimated from the lattice parameter of Pb, based on its equation of state\cite{36}. The program FULLPROF \cite{37} was used for the Rietveld refinement of both the XRD and NPD patterns, to determine the parameters associated with the crystal and magnetic structures of FGT.

First-principles calculations were performed on the basis of density-functional theory (DFT) using the generalized gradient approximation (GGA) in the form proposed by Perdew $\mathit{et}$ $\mathit{al}$. \cite{38}, as implemented in the Vienna $ab$ $initio$ Simulation Package (VASP) \cite{39,40}. The energy cutoff of the plane-wave was set to 500 eV. The energy convergence criterion in the self-consistent calculations was set to \textcolor{black}{10$^{-6}$} eV. A $\Gamma$-centered Monkhort-Pack $k$-point mesh with a resolution of 2$\pi$\texttimes{}0.03\textcolor{black}{{} \AA{}$^{-1}$} was used for the first Brillouin zone sampling. 

\section{\label{sec:level1} Results and discussions \protect\\}

Figure 1(a) shows the room-temperature XRD pattern of the synthesized polycrystalline sample, which can be well fitted by the expected hexagonal structure  of Fe$_3$GeTe$_2$ (space group: $P$6$_{3}$/$mmc$) alone without any additional reflections from impurities. The high purity of the sample is also confirmed by the ambient-pressure NPD pattern collected at room temperature, as shown in Fig. 2(a), which only consists of the reflections from FGT and Pb, the pressure marker.  By performing simultaneous refinements to the room-temperature XRD and NPD patterns, all structural parameters of FGT under ambient conditions are determined. As given in Table 1, the deficiency of the Fe(2) site ({$\delta$ = 5.0(1)$\%$}) in our polycrystalline sample is found to be quite minimal, accounting for its pretty high value of $T\rm_{C}$. Based on the dc magnetization data shown in Fig. 1(b), our polycrystalline sample exhibits a typical ferromagnetic (FM) behavior with the estimated Curie temperature as high as $T\rm_{C}$  = 228(1) K, which approaches the expected value for stoichiometric Fe$_3$GeTe$_2$ without any deficiency at the Fe(2) sites \cite{20} and is well consistent with the very small value of $\delta$.

\begin{center}
\includegraphics{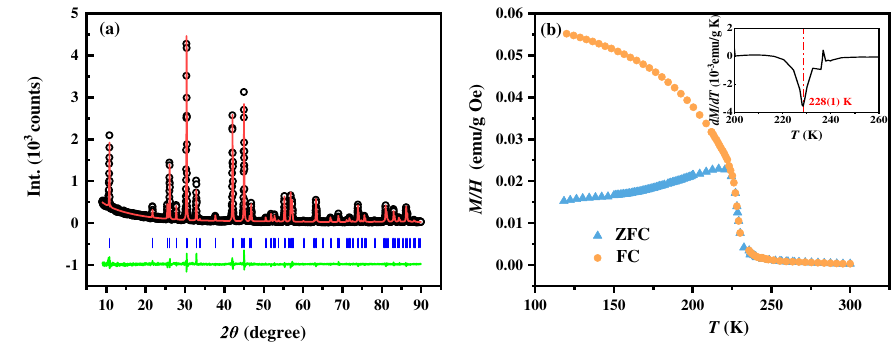}\\[5pt]  % insert figure
\parbox[c]{15.0cm}{\footnotesize{\bf Fig.~1.}   (a) Room-temperature XRD pattern of the polycrystalline FGT sample and the Rietveld refinement. The black open circles represent the observed intensities, and the red solid lines are the calculated patterns. The difference between the observed and calculated intensities is shown as the green solid line at the bottom. The blue vertical bars indicate the expected Bragg reflections from FGT. (b) DC magnetization of the polycrystalline FGT sample as a function of temperature, measured in an applied field of 100 Oe in zero-field-cooling (ZFC) and field-cooling (FC) modes, respectively. The inset shows the derivative (d$M$/d$T$) of the ZFC curve, where a clear dip is assigned as $T\rm_{C}$.}
\end{center}

\begin{center}
{\footnotesize{\bf Table 1.} Refinement results of the structural parameters of the polycrystalline FGT sample at ambient conditions.\\
\vspace{2mm}
\begin{tabular}{ccccccc}
\hline
{Atom} &{Wyckoff positions}   &{$x$}   &{$y$}   &{$z$}    &{$B\rm_{iso}$(\AA$^{2}$)} &{Occupancy}\\\hline
{Fe(1)} &{4$e$}	&{0} 	&{0} 	&{0.6725(2)}	&	{0.24(4)}	&1\\ 
{Fe(2)}	&{2$c$}	&0.6667	 	&0.3333 	&0.7500	&	0.24(4)	&0.95(1)\\ 
{Ge} &{2$d$}	&0.3333	 	&0.6667	 	&0.7500	&	0.6(1)	&1\\ 
{Te} &{4$f$}		&0.6667	 	&0.3333 	&0.5894(5)	&	0.04(9)	 &1\\\hline 
\multicolumn{7}{c}{$a$ = $b$ = 4.0307(1)\AA,  $     $ $c$ = 16.3575(6)\AA} \\  
\multicolumn{7}{c}{space group: $P$6$_{3}$/$mmc$} \\ 
\hline
\end{tabular}}
\end{center}

%The refinement yielded a=4.027(9)\AA and c=16.331(5)\AA, which is slightly different from the values reported in the literature obtained from room-temperature single-crystal x-ray diffraction, where a = 3.9910(10)\AA and c = 16.336(3)\AA. By the way, our refinement yields the partially occupied Fe\uppercase\expandafter{\romannumeral2} site , in agreement with both our NPD data and reference\cite{17}. According to the reference\cite{17}, the presence of different vacancies of Fe may result in variations in the lattice constant and Tc, which may explain this as well as the difference in P-T phase diagrm in Discussion. The Curie temperature (Tc) was determined using the temperature-dependent dc magnetic susceptibility $\chi$(T), as shown in Figure 1(b), and was found to be Tc = 228.5K.

Fig. 2(b) displays the ambient-pressure NPD pattern of FGT at 80 K, well below $T\rm_{C}$  = 228(1) K. Compared with the room-temperature pattern (Fig. 2(a)), no additional peaks can be identified at 80 K, but there clearly appears extra intensities on top of the (100) nuclear reflection as marked by the star, which is consistent with the expectation that it is in a FM state at 80 K with a magnetic propagation vector of $k$ = 0. The existence of two non-equivalent Fe sites in the unit cell and the overlap between the nuclear and magnetic reflections poses a challenge to our refinements to the NPD data. To reduce the strong correlations between individual refined parameters, we have fixed the ratio between the moment sizes at Fe(1) and Fe(2) sites as $M\rm_{Fe(1)}$/$M\rm_{Fe(2)}$ = 1.25, the value reported by Verchenko $et$ $al$. \cite{41}. As shown in Fig. 2(b), the ambient-pressure NPD pattern at 80 K agrees well with the reported FM structure of FGT with the magnetic moments at both Fe(1) and Fe(2) sites aligned along the $c$ direction (see Fig. 2(d)) \cite{17}, with the moment size of 1.5(1) and 1.2(1) $\mu\rm_{B}$, respectively. This magnetic structure agrees well with the magnetization data of the single-crystal samples of FGT \cite{42,43}, which confirms that the $c$ axis is its magnetic easy axis.

\begin{center}
\includegraphics{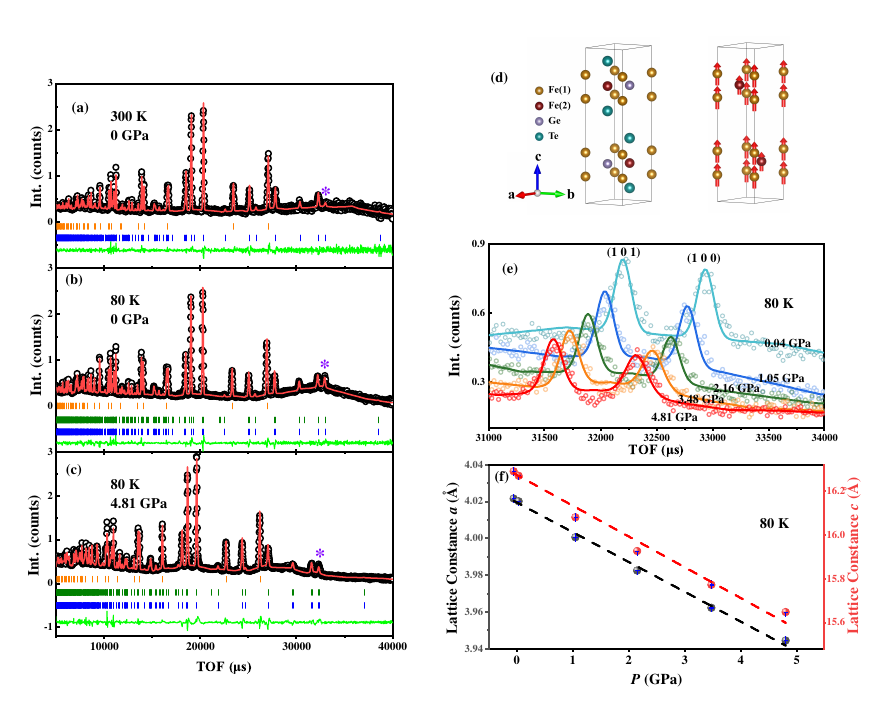}\\[5pt]  % insert figure
\parbox[c]{16.0cm}{\footnotesize{\bf Fig.~2.}  (a) and (b) show the ambient-pressure NPD patterns of FGT collected at room temperature and 80 K, respectively, while (c) shows the high-pressure NPD pattern of FGT collected at 80 K under $P$ = 4.809 GPa.  In (a, b, c), the black open circles represent the observed intensities, and the red solid lines are the calculated patterns. The differences between the observed and calculated intensities are shown as the green solid lines at the bottom. The blue, olive and orange vertical bars indicate the nuclear reflections from FGT, magnetic reflections from FGT, and nuclear reflections from Pb, respectively. The (100) reflection, where the $k$ = 0 magnetic scattering is strongest, is marked with the star. (d) illustrates the crystal and the FM structure of FGT. (e) shows an enlarged plot of the NPD patterns collected at 80 K under different applied pressures. (f) plots the determined lattice constants $a$ and $c$ at 80 K as functions of the pressure. The dashed lines are linear fittings.}
\label{fig:xrd}
\end{center}

At 80 K, as shown in Fig. 2(e), the nuclear Bragg peaks continously shift to the lower-TOF or higher-$Q$ side with increasing pressure, suggesting a dramatic shrinkage of the lattice constants $a$ and $c$ (see Fig. 2(f)) due to the effective compression generated by the pressure cell. No evidence of a pressure-driven structural phase transition is observed in our NPD data, indicating the stability of the hexagonal structure up to 4.809 GPa, which is consistent with the conclusion from previous high-pressure XRD measurements up to 25.9 GPa \cite{31}.  However, by compressing FGT up to $P$ = 4.809 GPa, as shown in Fig. 2(c), the additional intensities superimposed on the (100) nuclear refletion due to FM ordering decrease clearly, compared with the ambient-pressure case (Fig. 2(b)), suggesting a pressure-induced suppression of the ordered moments of Fe. 

Furthermore, sets of temperature-dependent NPD patterns were recorded under different pressures, to track the change of $T\rm_{C}$ with pressure. Fig. 3 summarizes the temperature dependences of the magnetic moment at the Fe(1) site determined from the refinements. It is clear that $T\rm_{C}$ is significantly suppressed from 225(5)K at the ambient pressure to 175(5)K at $P$ = 4.9(2) GPa, with a decay rate of $\sim$ 10 K/GPa. Through a linear extrapolation of the $T\rm_{C}$($P$) relation to higher pressures, a pressure-driven QCP is expected to emerge around $P\rm_{C}$ $\sim$ 22(2) GPa, as shown in the constructed $P$-$T$ phase diagram (Fig. 4). Although such a linear extrapolation is only a naive estimation and the actual $T\rm_{C}$($P$) relation is likely to deviate from it for the high-pressure region close to the QCP, the value of $P\rm_{C}$ estimated here based on the neutron diffraction study as a bulk and microscopic magnetic probe agrees nicely with the results from macroscopic anomalous Hall effect measurements under hydrostatic pressures \cite{31,32}, in which a decay rate of 9.2 K/GPa or 7.4 K/GPa is reported and a QCP around $P\rm_{C}$ = 21.2 GPa or 20.9 GPa is expected for the FGT sample with the $T\rm_{C}$ of 195 K or 155 K, respectively. Although the $T\rm_{C}$ values in different samples of FGT strongly depend on the Fe deficiency, their extrapolated $P\rm_{C}$ values tend to reach a convergence around 21-23 GPa, indicating the robustness of the FM order in FGT against the hydrostatic pressure. However, it is also worth noting that a paramagnetic ground state was speculated to emerge already above 15 GPa, based on a high-pressure synchrotron M{\"o}ssbauer source spectroscopic measurement on a FGT sample with $T\rm_{C}$ $\approx$ 220(5) K \cite{28}. Such a discrepancy might be due to that the synchrotron M{\"o}ssbauer source spectroscopy is actually a local magnetic probe with the beam size and penetration depth of the synchrotron x-ray limited to micrometers and may suffer from the inhomogeneity of the sample itself. Unfortunately, the maximal achievable pressure of $\sim$ 5 GPa with the MITO system at the PLANET beamline prevents a precise determination of $P\rm_{C}$ and further explorations into the quantum critical regime in FGT, in which the ferromagnetism of Fe will be fully suppressed and strong quantum fluctuations are expected to play an important role. 

\begin{center}
\includegraphics{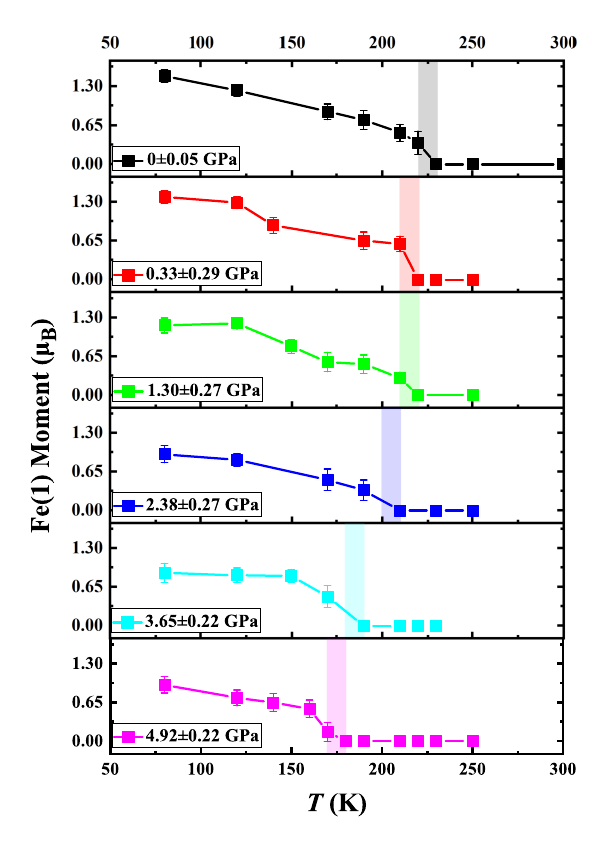}\\[5pt]  % insert figure
\parbox[c]{15.0cm}{\footnotesize{\bf Fig.~3.} Temperature dependences of the refined magnetic moment at the Fe(1) site, under different pressures ranging from $\sim$0 to $\sim$5 GPa, with the colored vertical bars corresponding to the $T\rm_{C}$. The lines are guides to the eye. The relatively large uncertainty of the pressure value in each panel is due to a slight release of the applied pressure during the cooling process.}
\end{center}

\begin{center}
\includegraphics{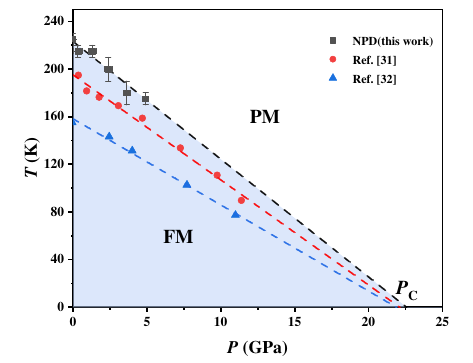}\\[5pt]  % insert figure
\parbox[c]{15.0cm}{\footnotesize{\bf Fig.~4.}  $P-T$ phase diagram of FGT constructed based on the $T\rm_{C}$($P$) relation determined from our NPD data. The data from high-pressure anomalous Hall effect measurements from Ref. [31] and [32] are also added. The dashed lines are linear extrapolations to the available data points as described in the main text.}
\end{center}

Based on the observed nearly linear dependence of $c$ and $a$ on the applied pressure up to $\sim$ 5 GPa, as shown in Fig.~2(d), we further assume linear evolutions of the lattice parameters for higher pressures and performed single-point DFT energy calculations at applied pressures of $0$, $5$, $15.8$, $25$, and $36.1~\mathrm{GPa}$, for the experimental determined $c$-axis aligned FM and paramagnetic (PM) structure, respectively. As shown in Fig.~5, the FM state is indeed energetically more stable below 25 GPa. With increasing pressure, the relative free energy of the PM state over the FM one gradually diminishes. On the other hand, the local magnetic moments of Fe(1) and Fe(2) ions also progressively decrease upon compressing and vanish at 36.1 GPa, as shown in Table 2. Our calculation here is well consistent with a previous DFT study, in which the calculated moments at both Fe(1) and Fe(2) sites are found to be finite at pressures up to 20 GPa \cite{26}. It seems that the DFT calculation inevitably overestimates the critical pressure $P\rm_{C}$, compared with Fig. 4, probably due to its intrinsic difficulty in dealing with the electronic correlation effects and the insufficiency of the local magnetism model for FGT, a vdW metal in which the itinerant electrons play an important role\cite{44,45}. However, these calculations qualitatively support the experimental conclusion that applying external pressure suppresses ferromagnetism towards a QCP.

    \begin{center}
\footnotesize{\bf Table 2.} DFT calculations of the magnetic properties of FGT under different applied pressure.\\
      \resizebox{\textwidth}{!}{
      \begin{tabular}{c|cc|cc|cc}
      \hline
      \multirow{2}[2]{*}{Pressure (GPa)} & \multicolumn{2}{|c|}{Lattice constant (\AA)} & \multicolumn{2}{|c|}{Local magnetic moment of FM ($\mu\rm_B$)} & \multicolumn{2}{c}{Relative Energy per Formula (meV)} \\
           & $a$    & $c$    & Fe(1) & Fe(2) & PM   & FM \\
      \hline
      0    & 4.03 & 16.35 & 2.473 & 1.567 & 952.06534 & 0 \\
      5    & 3.95 & 15.64 & 2.181 & 1.381 & 567.63616 & 0 \\
      15.8 & 3.764 & 14.05 & 1.465 & 1.113 & 150.254425 & 0 \\
      25   & 3.615 & 12.75 & 0.509 & 0.477 & 19.59723 & 0 \\
      36.1 & 3.435 & 11.2 & 0    & 0    & 0    & 0 \\
      \hline
      \end{tabular}}%
    \label{tab:DFT}
\end{center}

\begin{center}
    \includegraphics[width=0.6\textwidth]{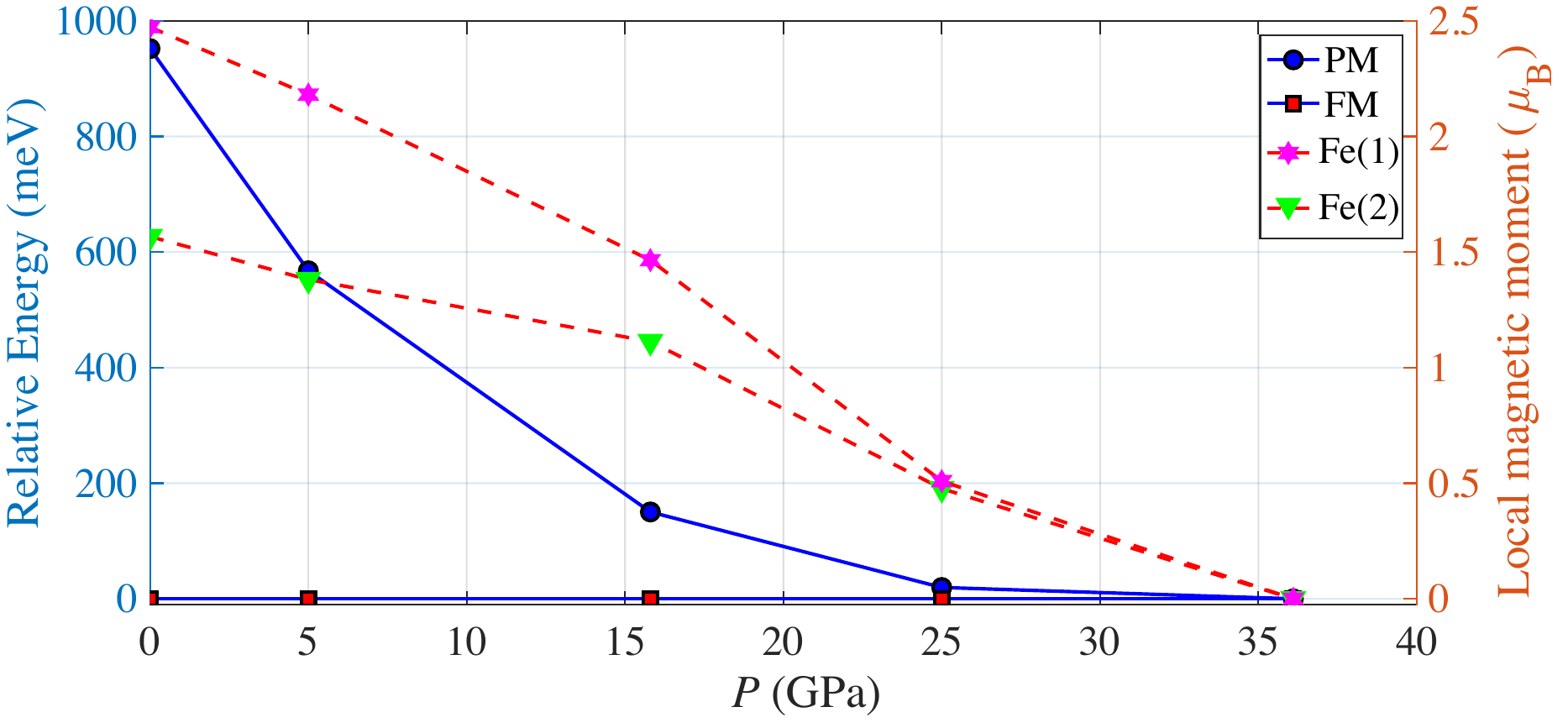}\\[5pt]  % insert figure
    \parbox[c]{15.0cm}{\footnotesize{\bf Fig.~5.}   Calculated relative energy of the PM state for FGT, compared with the FM one, and the calculated local magnetic moments for Fe(1) and Fe(2) ions, using the DFT method.}
    \label{fig:dft}
    \end{center}

The magnetism of FGT is largely due to the competition of various exchange interactions, including direct exchange interactions between the 3$d$ electrons of Fe and superexchange interactions through the Fe-Ge/Te-Fe paths. Due to the direct overlap of $d$ orbitals between nearest-neighbor Fe atoms, the resulting direct exchange interactions among the magnetic moments preferentially stabilizes antiferromagnetic (AFM) coupling \cite{46}. It can be found from Fig. 6(a) that both $d_1$ (the Fe(1)-Fe(1) bond length) and $d_2$ (the Fe(1)-Fe(2) bond length) shrink monotonically with increasing pressure, enhancing the AFM Fe-Fe coupling as the direct exchange interactions. In addition, the $d$ orbitals of Fe ions overlap with the $p$ orbitals of Te or Ge ions, facilitating virtual electron hoppings between the two nearest-neighbor Fe ions and leading to superexchange interactions, which are largely correlated with the angle between the atoms. According to the Goodenough-Kanamori-Anderson (GKA) rule, a 180° bond angle typically favors AFM coupling, while a 90° bond angle tends to incur FM couplings \cite{47,48,49}. As shown in Fig. 6(b), both $\theta_1$ (the Fe(1)-Te-Fe(1) bond angle) and $\theta_2$ (the Fe(1)-Ge-Fe(1) bond angle) progressively deviates from 90° as the pressure is increased, thus weakening the superexchange FM interactions. Together with the enhancement of the direct AFM interactions by pressure, the suppression of the ferromagnetism in FGT can be well understood.

\begin{center}
\includegraphics{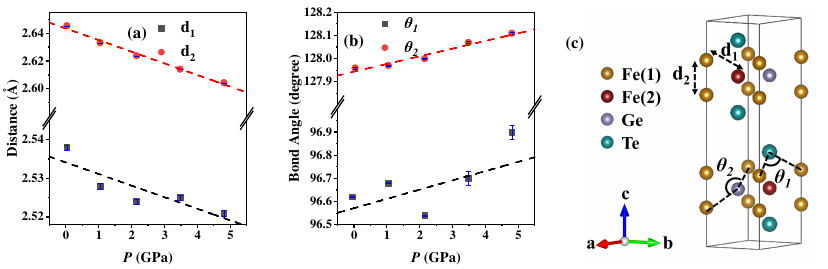}\\[5pt]  % insert figure
\parbox[c]{15.0cm}{\footnotesize{\bf Fig.~6.}  (a) and (b) show the pressure dependences of the $d_1$ and $d_2$ bond lengths, as well as the $\theta_1$ and $\theta_2$ bond angles, respectively, determined from the refinements to NPD patterns. The dashed lines are linear fittings to the data points. The definitions of $d_1$, $d_2$, $\theta_1$ and $\theta_2$ are illustrated in the unit cell shown in (c). }
\end{center}

\section{\label{sec:level1}Conclusion\protect\\}

In summary, almost stoichiometric FGT powders were synthesised by standard solid-state reaction method. High-pressure NPD measurement, as a microscopic and bulk magnetic probe, reveals the suppressive effect of hydrostatic pressure on the ferromagnetism of FGT, which is well supported by the DFT calculations. $T\rm_{C}$ decreases monotonically from 225(5) K to 175(5) K as pressure increases from 0 to 5 GPa, with a decay rate of $\sim$ 10 K/GPa. A pressure-driven QCP is expected at $P\rm_{C}$ = 22(2) GPa, implying the robustness of the ferromagnetism against the hydrostatic pressure. By performing Rietveld refinements to the NPD patterns, the changes of bond lengths and bond angles in FGT at different pressures were quantitatively determined. The application of hydrostatic pressure results in the continuous shrinkage of Fe-Fe bond lengths and progressive deviation of Fe(1)-Ge(Te)-Fe(1) bond angles from 90°, which significantly modifies the exchange interactions between the Fe ions and might account for the suppressive effect of hydrostatic pressure on the ferromagnetism.

\section*{Data availability statement}
The data that support the findings of this study are openly available in Science Data Bank at\\ https://www.doi.org/XXXXXXX. This statement should be given if some related data have been deposited in \href{https://www.scidb.cn/en}{Science Data Bank}.

\addcontentsline{toc}{chapter}{Acknowledgment}
\section*{Acknowledgment}
We acknowledge the financial support from the National Natural Science Foundation of China (Grant No. 12074023), the Large Scientific Facility Open Subject of Songshan Lake (KFKT2022B05), and the Fundamental Research Funds for the Central Universities in China. Neutron diffraction experiments at the Materials and Life Science Experimental Facility of the J-PARC were performed through the user program (Proposal No. 2023A0185).

%\end{CJK*}   end the Chinese environment
\end{document}